\documentclass[preprint,showpacs,preprintnumbers,amsmath,amssymb]{revtex4}

\usepackage{graphicx}
\usepackage{dcolumn}
\usepackage{bm}

\begin{document}

\title{Universal features and tail analysis of the order-parameter distribution of
the two-dimensional Ising model: An entropic sampling Monte Carlo study}

\author{Anastasios Malakis}
\altaffiliation{Corresponding author. Electronic address:
amalakis@phys.uoa.gr}
\author{Nikolaos G. Fytas}
\affiliation{Department of Physics, Section of Solid State
Physics, University of Athens, Panepistimiopolis, GR 15784
Zografos, Athens, Greece}
\date{\today}

\begin{abstract}
We present a numerical study of the order-parameter probability
density function (PDF) of the square Ising model for lattices with
linear sizes $L=80-140$. A recent efficient entropic sampling
scheme, combining the Wang-Landau and broad histogram methods and
based on the high-levels of the Wang-Landau process in dominant
energy subspaces is employed. We find that for large lattices
there exists a stable window of the scaled order-parameter in
which the full ansatz including the pre-exponential factor for the
tail regime of the universal PDF is well obeyed. This window is
used to estimate the equation of state exponent and to observe the
behavior of the universal constants implicit in the functional
form of the universal PDF. The probability densities are used to
estimate the universal Privman-Fisher coefficient and to
investigate whether one could obtain reliable estimates of the
universal constants controlling the asymptotic behavior of the
tail regime.
\end{abstract}

 \maketitle

\section{Introduction}
\label{section1}

A significant achievement in the theory of equilibrium critical
phenomena was the confirmation of universality and scaling
hypotheses and the calculation of critical
exponents~\cite{cardybook,fisher67,baxter,stanley,goldenfeld,kadanoff}.
The finite-size scaling technique has proven to be an extremely
reliable and powerful method for determining the critical
properties of low-dimensional systems, and several review articles
have appeared covering the original finite-size scaling theory and
later advancements~\cite{privman,fisher71,barber83,privman84}. In
practice all universality claims have been put to several tests
using data obtained from numerical simulations of finite systems
and the extraction of universal behavior from such studies remains
today of vigorous research interest. Of particular interest is the
view that, a strong hallmark of a universality class may be
obtained via the probability density functions of the main
thermodynamic variables of the system at
criticality~\cite{bruce85}. In finite systems the order-parameter
may be characterized by a probability distribution function (PDF)
and this distribution may be scaled appropriately to provide a
sequence of densities that appears to converge rapidly to a unique
scaling PDF~\cite{binder81,nicolaides88,patashinskii68}. We may
think of this limiting scaling function as the key ingredient
specific to a universality class, like the critical
exponents~\cite{bruce95}. This idea has been elaborated in recent
years by many authors, several papers are devoted to the
estimation of such universal distributions, and the subject is of
great and growing
interest~\cite{bruce85,bruce79,binder81,bruce95,nicolaides88,mon85,bruce92,hilfer95,smith95,hilfer03,botet00,eisenriegler87}.

Most properties of finite-size scaling functions are known from
numerical simulation of critical
systems~\cite{bruce79,binder81,patashinskii68,bruce95,nicolaides88,mon85,bruce92,hilfer95,smith95,hilfer03,botet00,eisenriegler87,cardy}.
Analytical results originate from field theoretic renormalization
group calculations~\cite{chen96,bruce81,brezin85}, conformal field
theory~\cite{cardybook,burkhardt85}, but also from a generalized
classification theory of phase
transitions~\cite{hilfer91,hilfer92,hilfer93,hilfer94}. In a few
cases some of the analytical predictions seem to have been
confirmed by numerical simulations~\cite{hilfer95,hilfer94}. A
detailed investigation of the tail regime was carried out in one
of the early multicanonical Monte Carlo simulations by Smith and
Bruce~\cite{smith95} for the two-dimensional Ising model, using
square lattices of size $L=32$ and $L=64$. Even though these
authors managed to measure extremely small tail probabilities with
high accuracy, and found signs for the far tail regime conjecture,
the theme was not fully elucidated. It is well-known that the
process of recording the very small probabilities in the tails of
the distribution requires specialized numerical techniques and
sufficient statistical accumulation is necessary in order to probe
the tails confidently.

Some studies using traditional Monte Carlo simulations have
attempted to estimate the universal density over the last years,
but failed in establishing the true behavior at the far tail
regime of the critical order-parameter distribution. Traditional
Monte Carlo sampling methods have increased dramatically our
understanding of the behavior of the standard classical
statistical mechanics systems. The Metropolis method and its
variants were, for many years, the main tools in condensed matter
physics, particularly for the study of critical
phenomena~\cite{metropolis53,bortz75,binder77,newman99,landau00}.
However, this standard approach has certain crucial weaknesses.
Importance sampling is trapped for very long times in valleys of
rough free energy landscape, as in complex systems in which
effective potentials have a complicated rugged landscape. For
large systems these trapping effects become more pronounced and
the Metropolis method, but also its variants, become inefficient.
Moreover, importance sampling methods are unsuccessful in
recording of the very small probabilities in the tails of the
critical order-parameter distribution. As was pointed out by
Hilfer \emph{et al.}~\cite{hilfer03} the study of the tail regime
requires special numerical
techniques~\cite{hilfer95,smith95,tsypin00}.

Several new efficient methods that directly calculate the density
of states (DOS) of classical statistical models play a dominant
role in overcoming the above mentioned problems in recent years. A
few remarkable examples of such methods are the
entropic~\cite{newman99,lee93}, multicanonical \cite{berg92},
broad histogram (BH)~\cite{oliveira96}, transition
matrix~\cite{wang02} and Wang-Landau (WL)~\cite{wang01} methods.
It is now possible to study more effectively the finite-size
scaling properties of statistical models by using entropic Monte
Carlo techniques. The multicanonical Monte Carlo method has
already been applied for the evaluation of the tail regime of the
universal PDF of the order-parameter~\cite{hilfer03} and the
present paper follows these attempts by using a new simple and
efficient entropic technique. This technique will be referred to
as the critical minimum energy subspace CrMES-WL entropic sampling
scheme~\cite{malakis05} and is in fact a program for the
simultaneous estimation of all thermal and magnetic finite-size
anomalies of the statistical system. The method was recently
presented and successfully tested on the square Ising
model~\cite{malakis05}. The central idea is to optimize the
simulations by using in a systematic way only the dominant energy
subspaces appropriate to the finite system at the temperature
range of interest. As shown in Ref.~\cite{malakis05} this
restriction speeds up our simulations, but also gives a new route
for critical exponent estimation by studying the finite-size
scaling of the extensions of the dominant subspaces. This scheme
is expected to be much more efficient than the Metropolis
algorithm, as already pointed out in a comparative
study~\cite{malakis05}, illustrating its superiority in the
recordings of the far tail regime. Overall, the aim of this
contribution is twofold: firstly to show that the CrMES scheme is
sufficient for the study of the tail regime and therefore to
propose a sensible optimization route. Secondly, to yield new
evidence confirming the tail regime conjecture and to provide an
early estimation of the universal parameters involved.

The rest of the paper is organized as follows. In the next
section, we briefly review the CrMES-WL entropic scheme, including
the N-fold way implementation of this scheme. This approach is
more efficient and makes available an additional approximation for
the DOS, based on the BH method. In Sec.~\ref{section3} we study
the order-parameter distribution of the square Ising model and we
verify its asymptotic behavior in the far tail regime. The
relevance of our findings to the universal Privman-Fisher
coefficient is also discussed and used to observe the
self-consistency of our estimations. Our conclusions are
summarized in Sec.~\ref{section4}.

\section{Entropic sampling in dominant energy subspaces}
\label{section2}

Let us briefly describe the Monte Carlo approach implemented in
order to generate the numerical data used in the next section.
Since the main ingredient of this approach, decisive for its
efficiency, is the restriction of the energy spectrum called CrMES
restriction, we shall also explain, although briefly, some key
ideas about its implementation. For a full description,
alternative definitions and further technical details of the CrMES
scheme one should consult the original papers, where the method
has been recently tested~\cite{malakis05,malakis04,martinos05}.
The main idea is to produce accurate estimates for all finite-size
thermal and magnetic anomalies of a statistical system by using a
DOS method, based on WL random walks, in an appropriately
restricted energy subspace $(E_{1},E_{2})$. In particular it was
shown~\cite{malakis05} that it is quite accurate to implement this
restricted scheme and at the same time accumulate data for the
two-parameter energy, magnetization ($E,M$) histograms, using for
their recording only the high-levels of the WL diffusion process.
At the end of the process the final accurate WL or the accumulated
BH approximation of the DOS and the cumulative ($E,M$) histograms,
are used to determine all properties of the statistical system.
This approximation is accurate in a wide temperature range,
depending on the extension of the dominant energy subspace used,
around the critical temperature of the system and this range can
be chosen in such a way that includes all pseudocritical
temperatures of the finite system.

A multi-range WL algorithm~\cite{wang01} is implemented to obtain
the DOS and the ($E,M$) histograms in the energy subspace
$(E_{1},E_{2})$. The WL modification factor $(f_{j})$ is reduced
at the jth iteration according to: $f_{1}=e,\; f_{j}\rightarrow
f_{j-1}^{1/2},\; j=2,...,J_{fin}$. The density of states obtained
throughout the WL iteration process may be denoted by $G_{WL}(E)$.
This density and the high-level ($j>12$, see also the discussion
below) WL ($E,M$) histograms (denoted by $H_{WL}(E,M)$) are used
to estimate the magnetic properties in a temperature range, which
is covered, by the restricted energy subspace $(E_{1},E_{2})$. The
N-fold version of the WL algorithm~\cite{schulz01,malakis04b}
makes available an additional approximation for the DOS (denoted
by $G_{BH}(E)$), based on the BH method. We shall use only the
approximation of $G_{BH}(E)$ corresponding to the minimum energy
transitions ($\Delta E=4$) in the N-fold process. $G_{BH}(E)$ is
obtained from the well-known BH equation~\cite{oliveira96}:
$G(E)\langle N(E,E+\Delta E)\rangle_{E}=G(E+\Delta E)\langle
N(E+\Delta E,E)\rangle_{E+\Delta E}$, where $N(E,E+\Delta E)$ is
the number of possible spin flip moves from a microstate of energy
$E$ to a microstate with energy $E+\Delta E$. The microcanonical
average of these numbers is estimated at the end of the CrMES-WL
entropic scheme, with the help of the corresponding appropriate
histograms recorded in the same high-level WL iteration range,
used also for the recording of the ($E,M$) histograms.

The probability density of the order-parameter at some temperature
of interest $T$ may be expressed as follows~\cite{malakis05}
\begin{subequations}
\label{eq:1}
\begin{equation}
\label{eq:1a}
P_{T}(M)\cong\frac{\sum_{E\in(E_{1},E_{2})}[H_{WL}(E,M)/H_{WL}(E)]G_{WL}(E)e^{-\beta
E}}{\sum_{E\in(E_{1},E_{2})}G_{WL}(E)e^{-\beta E}},
\end{equation}
where
\begin{equation}
\label{eq:1b} H_{WL}(E)=\sum_{M}H_{WL}(E,M)
\end{equation}
\end{subequations}
and the summation in $M$ runs over all values generated in the
restricted energy subspace $(E_{1},E_{2})$. Since the detailed
balance condition depends on the control parameter $f_{j}$, it is
suggested~\cite{malakis05} that only the high-level recipes or
their N-fold versions should be used for recording the ($E,M$)
histograms. This practice yields excellent estimates for all
magnetic properties as has been shown in detail in
Ref.~\cite{malakis05}.

The performance limitations of entropic methods, such as the WL
random walk and the reduction of their statistical fluctuations
have recently attracted considerable
interest~\cite{zhou05,dayal04,lee}. For the CrMES entropic scheme,
presented here in Eq.~(\ref{eq:1}), an extensive comparative study
using various implementations was presented in
Ref.~\cite{malakis05}. In particular, this study has clarified the
effect of the used range of the WL iteration levels on the
magnetic properties of the system and also the effect of one of
the simplest refinements of the WL algorithm. Statistical
fluctuations are reduced, as usually, by multiple measurements but
also by using the separation refinement proposed by Zhou and
Bhatt~\cite{zhou05}, in which the WL DOS modification is applied
after a number $\mathcal{S}$ ($\mathcal{S}=16$) spin-flips. For
the recordings of the ($E,M$) histograms, the range $j=12-24$, of
the WL iteration levels, was shown to give accurate estimates for
moderate lattice sizes ($L=10-120$), while a range of the order
$j=18-28$ would be a more safe choice for larger lattices.
Furthermore, as it has been shown
recently~\cite{malakis05,zhou05,lee}, the histogram flatness
criterion of the WL scheme for reducing the modification factor
should be treated with caution, and in order to avoid strong
statistical fluctuations in the final approximate DOS, enough
statistics should be obtained in each WL iteration.

Let us provide here some further technical details on the
implementation of our entropic scheme. For a large system, we
divide the total CrMES in several energy ranges of the order of
$50-60$ energy levels each, overlapping at their ends in at least
$3$ energy levels. When combining these energy intervals, simple
averaging is applied on the entropies of two neighboring pieces at
their overlapping end. Then, the DOS for each piece is adjusted to
the DOS of its previous neighbor and continuing this procedure we
finally obtain the DOS in the total CrMES. We follow a mixed WL
process in which in the first stage ($j=1-11$ or $j=1-17$) we use
the WL algorithm and in the second stage ($j=12-24$ or $j=18-28$)
its N-fold version. This mixed process was also used in a previous
study~\cite{malakis04b}, where a generalization of the N-fold
version was presented. In the present application only the
original Schulz \emph{et al.}~\cite{schulz01} N-fold version has
been used (case $c=1$ in Ref.~\cite{malakis04b}). The accumulation
of the microcanonical estimators, necessary for the application of
the BH equation~\cite{oliveira96}, takes place only in the N-fold
stage of the process. At these high levels of the WL random walk,
the incomplete detailed balanced condition has not a significant
effect on the BH microcanonical estimators and the particular
multi-range approach seems to be optimal in both time and accuracy
requirements. Besides the worse time requirements, our tests
indicated also more significant effects (presumably coming from
incomplete detailed balanced condition) for a multi-range approach
using much larger energy intervals. Furthermore, an accurate run
requires enough statistics to be obtained during each step of the
process.

The resulting WL and BH DOS approximations were tested using our
accurate finite-size data for the critical specific heat and the
asymptotic formulae discussed in
Refs.~\cite{malakis04,ferdinand69}. Using several independent
random walks, we have verified that both the final WL and BH
estimates for the critical specific heats ($C(T_{c})$) are very
accurate, with errors of less than $1\%$, even for large lattices
of the order of $L=200$. In many cases, the BH estimates showed
smaller statistical fluctuations when different runs were
compared. However, in cases in which insufficient simulations were
performed for large lattices ($L=140-200$), a significant
underestimation was observed for $C(T_{c})$ and this distortion
was then much stronger for the BH estimates. A further interesting
test, substantiating the statistical reliability of our approach,
was performed for the lattice $L=50$ by using the exact DOS
obtained by the algorithm provided by Beale~\cite{beale96}. In
this test, we applied a one-range exact entropic N-fold sampling
to obtain an approximate BH DOS and this was then used to obtain
an estimate $C(T_{c})$. When this one-range exact entropic
estimate of $C(T_{c})$ was compared with the corresponding
estimate obtained by our multi-range and approximate entropic
scheme, we found excellent coincidence and in effect the same
order of relative errors ($0.001 - 0.0001$) compared to the exact
result. Similarly, such small relative deviations, between the
exact and the approximate entropic schemes, were also observed in
the corresponding magnetic properties (for instance for
$\sqrt{\langle m^{2}\rangle}$) and in fact these deviations were
less significant than the fluctuations coming from limited
magnetic sampling, as observed in different runs. Therefore, we
suggest that the unified multi-range implementation of the
Wang-Landau~\cite{wang01} algorithm and the BH method of Oliveira
\emph{et al.}~\cite{oliveira96} may be advantageous in studies
using entropic schemes.

In the original paper~\cite{malakis04}, the CrMES method was
presented by restricting the energy spectrum around the value
$\widetilde{E}$, producing the maximum term in the partition
function at some temperature of interest. The restriction was
imposed by requesting a specified relative accuracy ($r$) on the
value of the specific heat. In other words, the restriction of the
energy spectrum produces an error on the value of the
thermodynamic parameter, at the particular temperature, and $r$
measures the produced relative error. This relative error is set
equal to a small number ($r=10^{-6}$), which is, in any case, much
smaller than the statistical errors of the Monte Carlo method (the
DOS method) used for the determination of the thermodynamic
parameter (for instance the specific heat). It was also shown that
a systematic study of the finite-size extensions of the resulting
dominant subspaces produces the thermal exponent $\alpha/\nu$ with
very good accuracy~\cite{malakis04}. As pointed out above, this
idea may be simultaneously applied~\cite{malakis05} for all
finite-size anomalies including the magnetic anomalies determined
from the ($E,M$) histograms. Furthermore, very good estimates of
the critical exponent $\gamma/\nu$ have been
obtained~\cite{malakis05} by studying the finite-size extensions
of analogous critical minimum magnetic subspaces (CrMMS) defined
also below. In this case, the restriction on the order-parameter
space around the value $\widetilde{M}$ that maximizes the
order-parameter density at the critical temperature $T_{c}$ is
imposed on the probability density function, as shown in
Eq.~(\ref{eq:2}). That is, the location of the dominant magnetic
subspaces ($\widetilde{M}_{-},\widetilde{M}_{+}$) may be obtained
by comparing the end-point densities with the peak-height of the
distribution
\begin{equation}
\label{eq:2} \widetilde{M}_{\pm}:\;\;\;
\frac{P_{T_{c}}(\widetilde{M}_{\pm})}{P_{T_{c}}(\widetilde{M})}\leq
r
\end{equation}
and the critical exponent $\gamma/\nu$ may be estimated from the
following scaling law ($\Delta
\widetilde{M}=(\widetilde{M}_{+}-\widetilde{M}_{-})$)~\cite{malakis05}
\begin{equation}
\label{eq:3}
\frac{(\Delta\widetilde{M})^{2}_{T_{c}}}{L^{d}}\approx
L^{\frac{\gamma}{\nu}}.
\end{equation}

One should note here that, the finite-size extensions of the above
defined CrMMS can be calculated by any Monte Carlo method
producing the order-parameter distribution. One could as well
implement the Metropolis algorithm to find estimates of the
extensions involved in Eq.~(\ref{eq:3}). However, this will yield
a marked underestimation of $\gamma/\nu$ as a result of the
statistical insufficiency of this traditional algorithm in the
tail regime. This effect was shown to be a result of the very slow
equilibration process of the algorithm in the far tail regime of
the order-parameter distribution~\cite{malakis05}. On the other
hand, the CrMES-WL entropic scheme was shown to produce very good
estimates of the critical exponent $\gamma/\nu$~\cite{malakis05}
and this can be taken as an indication in favor of the suitability
of this method in studies of the universal distributions and in
particular for their tail regime.

Let us now address the question of adequacy of the CrMES
restriction to deal with the far tail regime of the order
parameter PDF. In previous Monte Carlo
studies~\cite{hilfer03,smith95} extensive simulations were
undertaken in order to reach, as close as possible, the saturation
regime ($M/N\simeq 1,\; N=L^{2}$) of the order-parameter. This
practice is related to the fact that the part of the
order-parameter spectrum determining the tail regime is not
formally known. Scaling the PDF introduces a variable
$x=m/\sqrt{\langle m^{2} \rangle}$ and the (right) tail regime is
expected to be detected in the range
$x>1$~\cite{nicolaides88,smith95}. Consequently, a main question
concerns the sufficiency of the CrMES scheme. Does such a
restriction on the energy space yield a reliable approximation in
the tail regime? In particular, we would like to know whether this
restriction would allow us to simulate the order-parameter in the
appropriate range to confirm the tail regime conjecture. The
following observations provide strong evidence to this important
question and our findings in the next section establish explicitly
the fact that the implementation of a CrMES restriction permits an
asymptotic evaluation of properties related to the tail regime.

\begin{figure}[htbp]
\includegraphics*[width=12 cm]{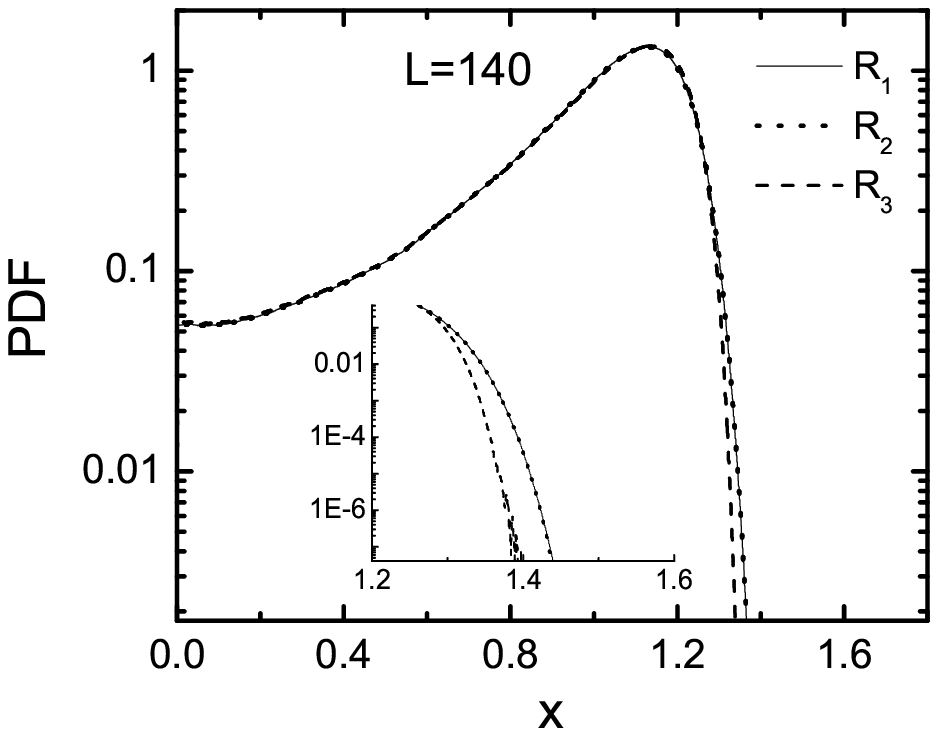}
\caption{\label{fig1}Universal PDF's for a lattice of linear size
$L=140$, with a logarithmic scale on the vertical axis. Three
different subspaces $R_{1},R_{2}$, and $R_{3}$ are used to obtain
the corresponding curves. The wider subspace $R_{1}$
($ie=1950-3800$) yields the PDF shown by the solid line. The dots
demonstrate the PDF corresponding to the subspace $R_{2}$
($ie=2228-3514$) and as can be seen from the inset the two cases
coincide in their common part. $R_{2}$ is the CrMES defined by the
specific heat's accuracy condition, as discussed in the text.
Finally, the dashed curve, $R_{3}$-case ($ie=2620-3800$),
illustrates that errors will be introduced by an inappropriate
restriction of the energy space.}
\end{figure}

In Fig.~\ref{fig1} we illustrate the effect of various
restrictions on the universal PDF (for details see next section)
for a large lattice size $L=140$. The simulation was carried out
in a wide energy range ($R_{1}$ in Fig.~\ref{fig1}) including the
CrMES at the exact critical temperature ($R_{2}$ in
Fig.~\ref{fig1}). Specifically, counting the energy levels from
the ground state with an integer variable $ie(=1,2,3,...)$, where
$ie=1$ corresponds to the ground state and the energy of a level
is $E=-2N+4\cdot(ie-1)$, the simulation was carried in the wide
subspace $R_{1}$: $ie=1950-3800$. The CrMES restriction applied to
produce a relative accuracy ($r=10^{-6}$) on the critical specific
heat (see the original paper on CrMES~\cite{malakis04}) yields the
subspace $R_{2}$: $ie=2228-3514$, while the restriction defined in
Eq.~(\ref{eq:2}) gives the subspace $ie=2246-3491$, which by
definition determines the extent we probe the tail of the
order-parameter distribution. We note that this later subspace is
included in the CrMES determined from the specific heat condition
(using the same level of accuracy $r$) and that this is a general
property which does not seem to depend on the lattice size, or
even on the model as far as we had the opportunity to observe.
This explains the striking observation from Fig.~\ref{fig1} that
the PDF's obtained from the subspaces $R_{1}$ and $R_{2}$
completely coincide in the $x$-range common in both subspaces. The
difference between the two curves is always smaller than the
accuracy level $r$ and, as one can observe more clearly from the
inset, the part of the $R_{1}$-PDF not accessible by using the
$R_{2}$(CrMES) subspace is the range: $x=1.47-1.59$. However, the
$R_{1}$-PDF appears to be already  flat in this range
$x=1.47-1.59$ and fitting attempts will be unstable, producing
large errors. Thus the saturation range is sensibly excluded from
our simulations by the CrMES restriction. The third curve, denoted
as $R_{3}$ in Fig.~\ref{fig1} corresponds to a PDF obtained in the
subspace $R_{3}$: $ie=2620-3800$, restricted severely from the
saturation side. This curve is presented in order to observe that
errors will be introduced if a restriction on the energy space is
inappropriately applied. Noting that the errors in determining the
end-points of the CrMES are of the order of $2-10$ energy levels
(at these lattice sizes) we conclude that critical minimum energy
subspaces, with a reasonable accuracy level of the order
$r=10^{-4}-10^{-6}$, will be sufficient for the study of the tail
regime.

\section{Universal features and the tail regime of the order-parameter distribution}
\label{section3}

\begin{figure}[htbp]
\includegraphics*[width=12 cm]{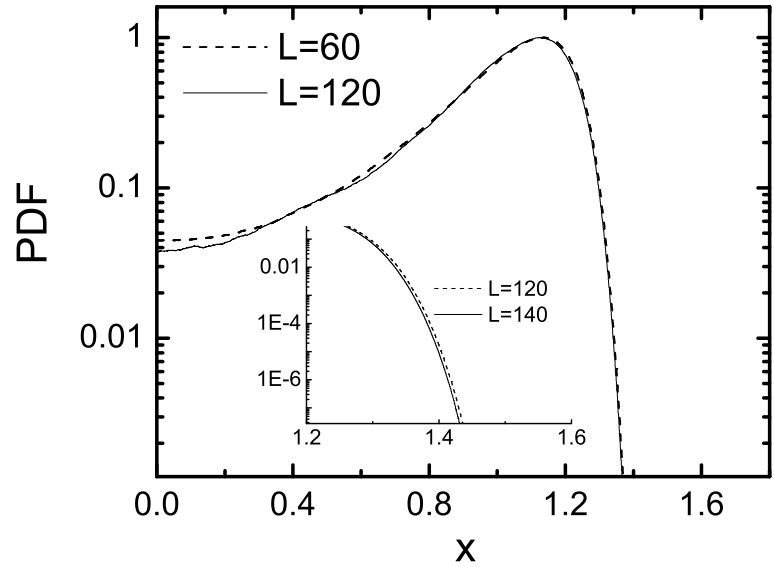}
\caption{\label{fig2}Illustration of the universal scaling
function $p(x)$ for $L=60$ and $L=120$. The inset is an
enlargement of the right tails for the case $L=120$ and $L=140$.
Note that the peak-heights have been set equal to $1$. In both
cases a logarithmic scale on the vertical axis has been taken.}
\end{figure}

The universal scaling distribution of the order-parameter may be
obtained from the magnetization distributions $p_{m}(m)$,
constructed with the help of the numerical scheme outlined in the
previous section, as follows~\cite{smith95}
\begin{equation}
\label{eq:4} p(x)dx\simeq p_{m}(m)dm,\;\; x=m/\sqrt{\langle m^{2}
\rangle},\;\; m=M/N.
\end{equation}
Figure~\ref{fig2} shows these distributions for lattice sizes
$L=60$ and $L=120$. The data used were generated by the
CrMES-WL(N-fold:14-24) entropic scheme using a separation
refinement $\mathcal{S}=16$, as explained in the previous section.
The curves shown in this figure illustrate the densities obtained
only via the BH approximation for the DOS and not the
corresponding WL approximation of Eq.~(\ref{eq:1}). This practice
will be followed bellow in all our figures, except for the cases,
indicated in the figures, where both the WL and BH DOS are used to
construct and illustrate the corresponding approximations for the
probability densities. Let us point out that the curves shown are
lines that pass through all the points representing the sampled
values of the order-parameter. For a large lattice, there will be
several thousands of such points, since each of them corresponds
to a possible value of $m$ ($=M/N,\;M=0,2,4,...,N$) of the finite
system. The density of points on the $x$-axis ($m$-axis) grows
with the lattice size, and we should expect to having
$N/2=L^{2}/2$ points in the positive $x$-axis ($m$-axis), provided
that all energies and all corresponding order-parameter values
were sampled by the WL process. However, as discussed in
Sec.~\ref{section2}, the points corresponding to the saturation
regime are not sampled, since this regime is excluded by the CrMES
restriction applied on the energy spectrum. In order to illustrate
the density functions in Fig.~\ref{fig2} we have chosen to
identify their peak-heights to the same value, set in
Fig.~\ref{fig2} equal to unity ($\hat{p}(x^{\ast})=1$, where
$x^{\ast}$ the most probable value). This is equivalent to
multiplying the universal PDF by a factor, which could in
principle be weakly $L$-dependent and this dependence will be
discussed further bellow.

\begin{figure}[htbp]
\includegraphics*[width=12 cm]{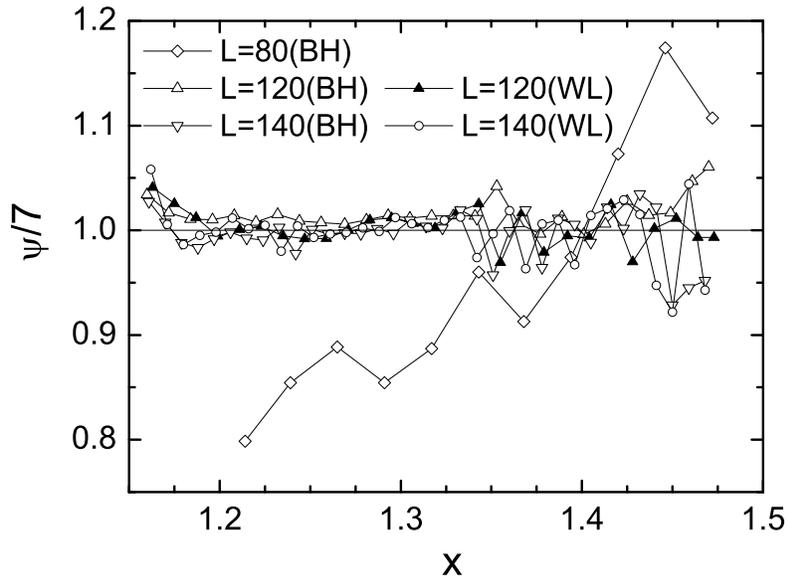}
\caption{\label{fig3}Behavior of estimates of the exponent
$\psi/7$ for lattice sizes $L=80,120$ and $140$, for both the WL
and BH methods used. The window $x=1.2-1.3$ appears to be very
stable for the sizes $L=120$ and $L=140$, where the estimates for
the exponent $\psi$ remain close to the value $\psi=7$.}
\end{figure}

Let us now consider the main conjecture for the large-$x$ behavior
of the universal function $p(x)$~\cite{smith95,hilfer95}
\begin{subequations}
\label{eq:5}
\begin{equation}
\label{eq:5a} p(x)\simeq
p_{\infty}x^{\psi}\exp(-a_{\infty}x^{\delta+1}),
\end{equation}
with
\begin{equation}
\label{eq:5b} \psi=\frac{\delta-1}{2}
\end{equation}
\end{subequations}
and $p_{\infty}$, $a_{\infty}$ universal constants. The structure
of the exponential (\ref{eq:5a}) has been suggested by rigorous
results for the two-dimensional Ising model~\cite{mccoy} and is
also consistent with Monte Carlo studies of the Ising universality
class~\cite{nicolaides88}. The studies of
Refs.~\cite{smith95,hilfer95} have provided some evidence for this
conjecture and in particular for the prefactor and the relation of
the exponent $\psi$ to the critical exponent $\delta$. For the 2D
Ising model the exponent should have the value $\psi=7$, if, of
course, the prefactor hypothesis is valid. Smith and
Bruce~\cite{smith95} provided numerical support for this value,
but their study was not completely conclusive since it was carried
out only for relatively small lattices ($L=32$ and $L=64$) and the
$x$-window in which the value $\psi=7$ was observed was actually
quite narrow. We now present results for several lattice sizes
($L=80,100,120$, and $140$) reinforcing this conjecture in a very
wide $x$-window. Following Smith and Bruce~\cite{smith95} we fix
the exponent $\delta$ in the exponential factor of
Eq.~(\ref{eq:5}) and fit our results $(x>1)$ in $x$-windows, each
one corresponding to $50$ different magnetization values, sampled
during the WL(N-fold:12-24) process. Fig.~\ref{fig3} shows a very
clear signature of the prefactor law (\ref{eq:5b}) which upholds
in a large $x$-window only for the large lattices ($L=120$ and
$L=140$ are shown). On the other hand, for smaller lattice sizes
$L\lesssim 80$ ($L=80$ is shown) the picture is similar to that
presented in Ref.~\cite{smith95} and the expected value is
obtained only in a small $x$-window. In Fig.~\ref{fig3} we have
illustrated the behavior of the estimates for the exponent $\psi$,
and for the large lattices ($L=120$ and $L=140$) both the WL and
BH cases are shown. From this figure we observe that the window
$x=1.2-1.3$ is very stable for the sizes $L=120$ and $L=140$, and
the estimates for the exponent $\psi$ remain close to the value
$\psi=7$, beyond small fluctuations, for even larger values of
$x$. It appears that this stable $x$-window may be very convenient
for further fitting attempts.

\begin{figure}[htbp]
\includegraphics*[width=12 cm]{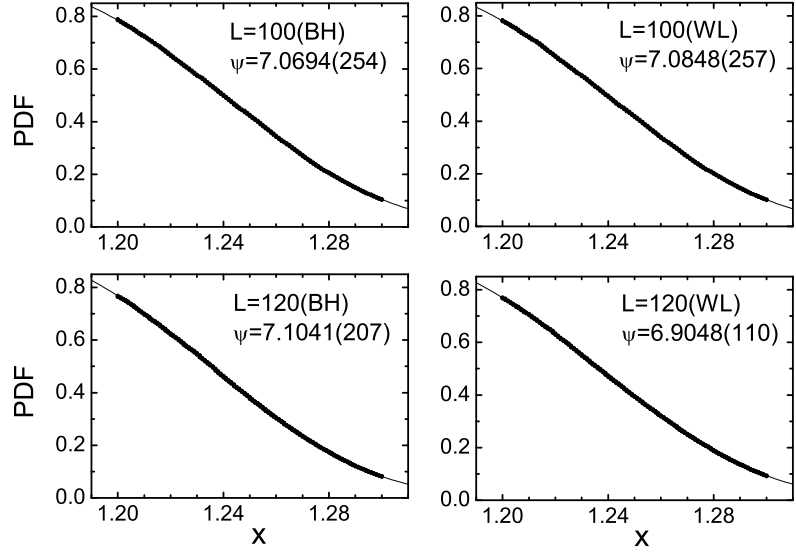}
\caption{\label{fig4}Illustration of the three-parameter fitting
attempts for $L=100$ and $L=120$ using both WL and BH
approximations of the universal PDF. The values of the estimates
for the exponent $\psi$ and their fitting errors manifest the
accuracy of our results and also the stability of the fitting
attempts at the window $x=1.2-1.3$. The small differences of the
exponent estimates from the two different DOS's, within the same
entropic sampling runs, reflect the sensitivity of the fitting
attempts to statistical errors, growing with the lattice size.}
\end{figure}

Encouraged from the above finding, we now use this stable
$x$-window to perform further fitting attempts, allowing this time
both exponents in the exponential and in the prefactor term vary,
and use $\psi$ as a free parameter by assuming the validity of
Eq.~(\ref{eq:5b}). In this way we examine whether the particular
$x$-window provides a good independent estimation of the exponent
$\delta$, based on the full tail regime conjecture
[Eq.~(\ref{eq:5})]. In other words, $\psi$, $a_{\infty}$ and
$p_{\infty}$ are the three free fitting parameters in applying
Eq.~(\ref{eq:5}) to our numerical data. Figure~\ref{fig4} depicts
the three-parameter fitting attempts for $L=100$ and $L=120$ using
both WL and BH approximations of the universal PDF. The values of
the estimates for the exponent $\psi$ and their fitting errors are
illustrated in this figure. The estimates obtained for the
exponent $\psi$ are quite good in all cases and their average
$\bar{\psi}=7.041$ is accurate to two significant figures, which
is a rather pleasing result.

It is interesting to examine whether one could obtain reliable
estimates of the universal constants $a_{\infty}$ and $p_{\infty}$
using the above window and to see whether such estimates can be
tested against known results in the literature. It appears that
the idea of such a test has not been tried in previous studies. In
contrast, the main conclusion of the recent paper of Hilfer
\emph{et al.}~\cite{hilfer03} is that ``the universal scaling
function for the order-parameter distribution cannot be considered
to be known from numerical simulations at present''. Furthermore,
it was stated in the conclusion of this paper that ``even for the
square Ising model a numerical estimation of the universal PDF
requires sizes that are beyond present day computer recourses''.
Therefore, one would be tempted to think that, also the universal
constants $a_{\infty}$ and $p_{\infty}$ will approach their values
slowly in the asymptotic limit. Thus, a reliability test for the
values of $a_{\infty}$ and $p_{\infty}$ estimated in the above
$x$-window  will be useful even in the case where it fails. On the
other hand, a test yielding a good comparison will be a strong
verification of the proposal that the observed stable $x$-window
can be used for the extraction of the asymptotic behavior of the
tail regime of the universal scaling function.

In order to construct a reliability test as discussed above, we
shall now define following Bruce~\cite{bruce95} the universal
function $\mathcal{F}(y)$
\begin{equation}
\label{eq:6} \mathcal{F}(y)=\ln\left[\int dx\;p(x) e^{yx}\right].
\end{equation}
The limiting behavior of the function $\mathcal{F}(y)$ is
controlled by the large-$x$ behavior of $p(x)$ and as shown by
Bruce~\cite{bruce95} one can find, by assuming the validity of the
full ansatz (\ref{eq:5}), a large $y$-expansion of
$\mathcal{F}(y)$ which reads as
\begin{equation}
\label{eq:7} \mathcal{F}(y)\simeq
b_{\infty}y^{1+1/\delta}+\frac{1}{2}\ln\left[\frac{2\pi
p_{\infty}^{2}}{a_{\infty}\delta (\delta+1)}\right],
\end{equation}
where $b_{\infty}$ is a constant. The constant term in
Eq.~(\ref{eq:7}) is the universal Privman-Fisher
coefficient~\cite{bruce95}. Its value for the 2D square Ising
model is $U_{o}=-0.639\;912$~\cite{bruce95,ferdinand69}, and as
shown in Ref.~\cite{bruce95} is related via the
ansatz~(\ref{eq:5}) to the universal constants $a_{\infty}$ and
$p_{\infty}$ as follows
\begin{equation}
\label{eq:8} U_{o}=\frac{1}{2}\ln\left[\frac{2\pi
p_{\infty}^{2}}{a_{\infty}\delta (\delta+1)}\right].
\end{equation}
The expansion (\ref{eq:7}) above has been fully illustrated by
Bruce~\cite{bruce95}, using appropriate $y$-windows to observe the
development of the effective $U_{o}$ ($U_{o}^{eff}$) in the large
$y$-range. The agreement of this development towards the universal
Privman-Fisher coefficient was shown to be excellent (see Fig. 1
in Ref.~\cite{bruce95}).

\begin{figure}[htbp]
\includegraphics*[width=12 cm]{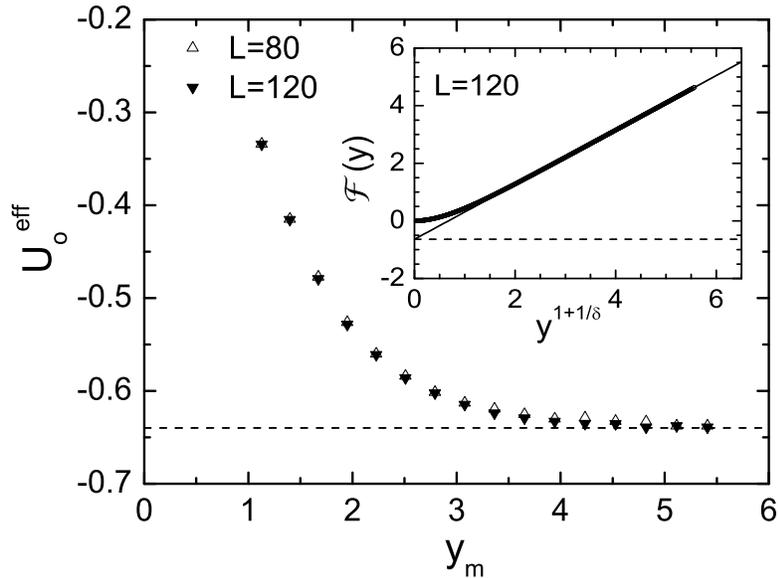}
\caption{\label{fig5}Variation of $U_{o}^{eff}$, for lattices
$L=80$ and $L=120$, obtained as intercepts of linear fittings of
Eq.~(\ref{eq:7}) in small windows in the variable
$y^{1+1/\delta}$. The $y_{m}$ values correspond to the center of
the fitting windows. The inset shows the linear fit according to
Eq.~(\ref{eq:7}) in the large $y$-part ($L=120$). The dashed line
marks the exact value of the amplitude $U_{o}$ for the 2D Ising
model~\cite{ferdinand69}, approached by the corresponding
intercepts. This figure should be compared to Fig. 1 of
Ref.~\cite{bruce95}.}
\end{figure}

In  Fig.~\ref{fig5} we reproduce Fig. 1 of Ref.~\cite{bruce95},
using our numerical data for sizes $L=80$ and $L=120$. The
approach to the universal value $U_{o}$ is again excellent and the
$L=120$ data produce a slightly faster approach, as can be seen
from this figure. The estimates of $U_{o}^{eff}$ in the range of
$y_{m}\simeq 5$ deviate from the exact value by less than $0.2\%$.
This is an explicit confirmation that a CrMES restriction is
adequate for (numerical) studies of the tail regime. As pointed
out by Bruce~\cite{bruce95}, the numerical estimation of $U_{o}$
via the numerical integration (see Eq.~(\ref{eq:6})), illustrated
above, is more reliable than attempting a direct Monte Carlo
determination of free energies~\cite{bruce95,mon85} or attempting
to estimate $U_{o}$ via the factors appearing in Eq.~(\ref{eq:8}).
In accordance with our earlier discussion, an accurate estimation
of the universal constants $a_{\infty}$ and $p_\infty$ would not
be normally expected at these lattice sizes and corrections to
scaling may complicate the situation.

We proceed to describe an estimation procedure for the universal
constants $a_{\infty}$ and $p_{\infty}$, based on the stable
window $x=1.2-1.3$. As can be seen from the fitting attempts in
Fig.~\ref{fig4}, the estimates for the critical exponent $\psi$
suffer from statistical fluctuations and are sensitive to both the
DOS statistical method (WL or BH) and to the lattice size. It is
therefore very important to repeat the statistical sampling and to
use several lattice sizes in order to be able to observe
systematic dependencies that may effect the estimated values. A
fully converged implementation of the WL algorithm will be
essential for the accurate application of the scheme and, of
course, as usually repeated applications should improve the
scheme.

Thus, we have applied a well-saturated (see the discussion in
Sec.~\ref{section2}) CRMES-WL(N-fold:16-30) entropic scheme using
a separation refinement $\mathcal{S}=8$ for the N-fold levels
$j=16-26$ and a separation refinement $\mathcal{S}=16$ for the
N-fold levels $j=27-30$. In each run we used 4 such WL random
walks to estimate the average WL DOS and to obtain the
corresponding BH DOS. At the end of the run the cumulative
$H_{WL}(E,M)$ histograms were used with the above DOS's to obtain
the corresponding universal PDF's. This process was repeated $4$
times for each of the lattices sizes $L=90,100,110,120$ and $140$.
For each lattice size, and each of the above runs (consisting of
the $4$ WL random walks) we fitted the function (\ref{eq:5a}) in
the stable window $x=1.2-1.3$ by fixing the value of the exponent
$\psi$ to the expected value $\psi=7$. Thus, for each lattice size
we obtained mean values and statistical errors for the following
parameters: the universal constant $a_{\infty}$ and
$\hat{p}_{\infty}$, where $\hat{p}_{\infty}$ absorbs the
identification of all peak-heights to $1$. The multiplying factor,
thus absorbed in our notation by using $\hat{p}_{\infty}$, is the
peak-height $p(x^{\ast})$ ($p_{\infty}=\hat{p}_{\infty}\cdot
p(x^{\ast})$). The peak-height was also estimated for each run
separately as well as the parameter
$\lambda=L^{-1/8}/\sqrt{\langle m^{2} \rangle}$ which connects the
universal PDF in the scaling variable $x=m/\sqrt{\langle m^{2}
\rangle}$~\cite{bruce95} with the universal PDF in the scaling
variable $z=mL^{1/8}$ used by Hilfer \emph{et al.}~\cite{hilfer95}
($z=\lambda^{-1} x$). Our fitting attempts were also repeated in
this $z$-representation using the equivalent stable window which
has been taken to be $z=1.26-1.36$. For the parameters in this
$z$-representation the following notation was used: $p_{z,\infty}$
and $a_{z,\infty}$ for the universal constants controlling the
tail regime, and $p_{z}(x^{\ast})$ for the peak-height.

Using the above described fitting attempts we observed no
systematic $L$-dependency, in the range $L=90-140$. This
observation concerns all the estimated parameters, and it appears
that not even weak $L$-corrections should be applied. In fact the
picture, obtained for each parameter, resembles effectively a
statistical fluctuation around a mean value. Therefore it seems
sensible to average the values of all parameters over the range
$L=90-140$ (assuming no systematic $L$-dependency) and take as
respective errors three standard deviations of the averaging
process. The values of all parameters obtained from this
hypothesis are collected in Table~\ref{table1} and are presented
for the methods corresponding to: the WL DOS (WL) and the BH DOS
(BH). Both $x$- and $z$-representations were used in the fittings
and the average for $\lambda$ is also given in the footnote of
Table~\ref{table1}. Using these values in conjunction with
Eq.~(\ref{eq:8}) and the exact value for the state exponent
$\delta=15$ (note also that $p_{\infty}=\hat{p}_{\infty}\cdot
p(x^{\ast})$), the respective estimate for the universal
Privman-Fisher coefficient is determined for each case and is
presented in Table~\ref{table1}.

Discussing Table~\ref{table1} we first observe that all parameters
involved in the estimation of the universal Privman-Fisher
coefficient have reasonable relative errors of the order of
$2-4\%$. The corresponding relative errors for $U_{o}$ are of the
order of $4-10\%$. However, the values obtained for $U_{o}$ are
lower than the exact value and the underestimation is of the order
of $7\%$. This, compared to the excellent agreement obtained by
the numerical integration method shown in Fig.~\ref{fig5} and
discussed above reveals the superiority of the integration method.
The origin of the present underestimation is not clear. One may
think that, moving to a $x$-window corresponding to larger values
of $x$ could improve the above picture. Nevertheless, we have
checked that this is not so, and that statistical errors become
dominant as we move to larger $x$-values, making the estimation
scheme unreliable. Finally, let us compare the constant
$a_{z,\infty}=0.026...$ of Table~\ref{table1} with the coefficient
of the exponential term given in Eq. (7) of Ref.~\cite{hilfer95}.
This latter coefficient has the value
$\frac{1}{\delta}(\frac{\delta}{\delta+1})^{\delta+1}=0.023738...$
for the square Ising model. These two values are of the same
order, although corresponding to different scaling
limits~\cite{hilfer95}, and this is an indication in favor of the
reliability of the stable $x$-window discussed here. It also
strongly supports the proposal of Hilfer and
Wilding~\cite{hilfer95} to calculate critical finite-size scaling
functions via the generalized classification theory of phase
transitions developed by
Hilfer~\cite{hilfer91,hilfer92,hilfer93,hilfer94}. It will be
interesting to uncover the reasons behind the minor
underestimation of the universal Privman-Fisher coefficient by the
direct estimation of the universal constants $a_{\infty}$ and
$p_{\infty}$, as attempted in this work.

Finally, we briefly discuss the behavior of the left tails of the
critical distributions. This behavior has been considered by
Hilfer \emph{et al.}~\cite{hilfer03} by using rescaled probability
densities $p_{o}(x)$, where $x$ and $p_{o}$ are defined in such a
way that yield mean zero, unit norm, and unit
variance~\cite{hilfer03}. This new scaled variable $x$ should not
be confused with the scaled variable $x$ defined in
Eq.~(\ref{eq:4}) and appearing in Figs.~\ref{fig1} - \ref{fig4}.
These new distributions are appropriately translated with respect
to the position of the peak of the distribution ($x_{peak}$) for
each tail separately. For instance for the left tail one defines:
$p_{ol}(x)=p_{o}(x_{peak}-x)$ with $x<x_{peak}$. Furthermore,
these authors used plots of the functions $q(y)=\frac{d\log(-\log
p_{ol})}{d\log x}$ versus $y=\log x$, and compared their tail
behavior with the standard Gaussian behavior. The large $x$-range
of these plots is convenient for illustrating the possible
developments of fat stretched exponential tails of the form $\sim
\exp[-x^{\omega}]$. Using their method we have also tried to
clarify the behavior of the left tails of our critical
distributions for all sizes up to $L=140$. Qualitatively, we found
the same behavior with that of Ref.~\cite{hilfer03}. In particular
a data collapse for all sizes up to $L=140$ was observed and the
picture is very similar with their behavior (see: the right upper
row of Fig. 6 in Ref.~\cite{hilfer03}). However, it should be
noted that, the data corresponding to the left tails suffer from
relatively stronger fluctuations when compared with the data of
the right tails. This is due to the inevitable more limited
magnetic sampling in the left tails. Note that, the left tails are
strongly influenced from a part of the energy spectrum in which a
much larger number of spin configurations exists, compared with
the lower energy part of the spectrum determining the right tails.
However, using relatively larger $x$-windows (corresponding to at
least $250$ different magnetization values for $L=120$) we have
tried to repeat the small $x$-windows fitting attempts, as those
shown in our Fig.~\ref{fig3}. In this case we used as a test
function the simple exponential behavior $p_{ol}(x)=q\exp[-b
x^{\omega}]$, and the total fitting range of the new scaled
variable $x$ was $x=1-3.25$, which we assume describes the left
tail regime. Note that this range corresponds to the range
$x=0.87-0.28$ of our Fig.~\ref{fig2} for the scaled variable
defined in Eq.~(\ref{eq:4}). For the lattice $L=120$ the total
fitting range includes about $2400$ different magnetization
values. Remarkably, we have obtained almost the same behavior for
$L=50$ and $L=120$. For both sizes we found that for the above
rather large range ($1-3.25$) the exponent $\omega$ steadily
decreases (almost linearly for $L=120$) from the value
$\omega=0.75$ to the value $0.15$ passing through the value
$\omega=0.5$ in the neighborhood of $x=1.75$. For larger values of
$x$ outside the above range ($1-3.25$) the statistical
fluctuations do not permit reasonable fits. The described behavior
is consistent with that of Ref.~\cite{hilfer03} and since it is
observed for the small and the larger lattice sizes we may suppose
that it represents also the true asymptotic behavior. As a final
remark we point out that the fat stretched exponential left tail
$\sim \exp[-\sqrt{x}]$ reported in Ref.~\cite{hilfer03} at the low
temperature $T=1.5$ can not be studied by our data. Such a study
would require the application of the present scheme in a lower
part of the energy space.

\begin{table*}
\caption{\label{table1}Estimates for the universal parameters and
the Privman-Fisher coefficient.}
\begin{ruledtabular}
\begin{tabular}{cccccc}
Method\footnotemark[1] & &   & & & $U_{o}$ \\
\hline\hline &$\hat{p}_{\infty}$ &  $p(x^{\ast})$ &$p_{\infty}$ &$a_{\infty}$ &  \\
WL &0.6258(240)&  1.309(51)&0.8190(449)& 0.0571(32) & -0.589(63)\\
BH &0.6172(110)&  1.303(30)&0.8042(234)& 0.0561(20) & -0.599(34)\\
\hline\hline &$\hat{p}_{z,\infty}$ &  $p_{z}(x^{\ast})$ &$p_{z,\infty}$ &$a_{z,\infty}$ &  \\
WL &0.4476(269)& 1.252(52)&0.5604(156)& 0.0266(9) & -0.589(33) \\
BH &0.4462(132)& 1.246(30)&0.5560(213)& 0.0268(5) & -0.599(40)
\end{tabular}
\end{ruledtabular}
\footnotetext[1]{WL: $\lambda=0.951(11)$, BH: $\lambda=0.953(5)$.}
\end{table*}

\section{Conclusions}
\label{section4}

In the present Monte Carlo study we applied the CrMES-WL entropic
sampling scheme, based on the high-levels of the Wang-Landau
process in dominant energy subspaces, in order to obtain the
order-parameter universal PDF of the square Ising model for large
lattices. The efficiency of this method enabled us to clarify the
asymptotic tail behavior of the universal distribution and to
obtain reliable data for the universal parameters. In particular,
we found that there exists a large stable window of the scaled
order parameter in which the full ansatz for the tail regime is
well obeyed. In a second stage, this window was used to estimate
the equation of state exponent $\delta$ and also to observe the
behavior of the universal constants implicit in the functional
form of the universal PDF and to approximate for the first time
their values. The estimates of the universal constants appear to
be reliable to within $3-7\%$ of statistical errors. The excellent
accuracy obtained for the universal Privman-Fisher coefficient, by
appropriate numerical integration, was also illustrated and
consist a concrete reliability test of the accuracy of our
numerical scheme.

\begin{acknowledgments}
{This research was supported by EPEAEK/PYTHAGORAS under Grant No.
$70/3/7357$. One of us (N.G.F.) would like to thank the
Empeirikeion Foundation for financial support.}
\end{acknowledgments}

{}

\end{document}